\documentclass[longauth,utf8]{aa}
\usepackage[varg]{txfonts}
\usepackage{natbib}
\usepackage{xspace}
\usepackage{url}
\usepackage{booktabs}
\usepackage{footmisc}
\usepackage{graphicx}
\usepackage{amssymb, amsmath}
\usepackage{color}
\usepackage{bm}
\bibpunct{(}{)}{;}{a}{}{,} 

\usepackage{xspace}
\usepackage{siunitx}
\usepackage[export]{adjustbox}
\usepackage{lineno}
\usepackage{hyperref}

\makeatletter
\renewcommand*\aa@pageof{, page \thepage{} of \pageref*{LastPage}}
\makeatother

\usepackage{bm}
\newcommand{\vect}[1]{\boldsymbol{\mathbf{#1}}}

\DeclareSIUnit\year{yr}
\DeclareSIUnit\erg{erg}
\DeclareSIUnit\au{AU}
\DeclareSIUnit{\msun}{\mbox{$M_{\odot}$}}
\DeclareSIUnit{\rsun}{\mbox{$R_{\odot}$}}
\DeclareSIUnit\ev{eV}
\DeclareSIUnit\kev{\kilo\ev}
\DeclareSIUnit\mev{\mega\ev}
\DeclareSIUnit\gev{\giga\ev}
\DeclareSIUnit\tev{\tera\ev}
\DeclareSIUnit\pe{p.e.}
\DeclareSIUnit\parsec{pc}
\DeclareSIUnit\gauss{G}
\DeclareSIUnit\lightyear{ly}

\def \deg{^\circ}

\def \gammapy{\texttt{gammapy}\xspace}
\def \fermi{\textit{Fermi}-LAT\xspace}
\def \magic {MAGIC\xspace}
\def \veritas{VERITAS\xspace}
\def \fact{FACT\xspace}
\def \hess{H.E.S.S.\xspace}
\def \cta{{CTA}\xspace}
\def \hessOne{H.E.S.S. Phase I\xspace}
\def \hessTwo{H.E.S.S. Phase II\xspace}

\def \gammapyUrl{\url{https://gammapy.org/}\xspace}

\def \gadfUrl{\url{https://gamma-astro-data-formats.readthedocs.io}\xspace}

\def \hessDataUrl{\url{https://www.mpi-hd.mpg.de/hfm/HESS/pages/dl3-dr1/}}

\def \apjl {ApJL}
\def \apjs {ApJS}

\makeatletter
\renewcommand*{\@fnsymbol}[1]{\ifcase#1\or*\or$\dagger$\or$\ddagger$\or**\or$\dagger\dagger$\or$\ddagger\ddagger$\fi}
\makeatother

\begin{document}
\title{Towards open and reproducible multi-instrument
analysis\\ in gamma-ray astronomy}
\titlerunning{Multi-instrument gamma-ray astronomy}

\author{C.~Nigro \inst{\ref{DESY}}$^{\star}$ 
\and C.~Deil \inst{\ref{MPIK}}
\and R.~Zanin \inst{\ref{MPIK}}
\and T.~Hassan \inst{\ref{DESY}}
\and J.~King \inst{\ref{HDL}}
\and J.E.~Ruiz \inst{\ref{IAA}}
\and L.~Saha \inst{\ref{UCM}}
\and R.~Terrier \inst{\ref{APC}}
\and K.~Br{\"u}gge \inst{\ref{DORTMUND}}
\and M.~N{\"o}the \inst{\ref{DORTMUND}}
\and R.~Bird \inst{\ref{UCLA}}
\and T.~T.~Y.~Lin \inst{\ref{MCGILL}}
\and J.~Aleksi\'c \inst{\ref{IFAE}}
\and C.~Boisson \inst{\ref{LUTH}}
\and J.L.~Contreras \inst{\ref{UCM}}
\and A.~Donath \inst{\ref{MPIK}}
\and L.~Jouvin \inst{\ref{IFAE}}
\and N.~Kelley-Hoskins \inst{\ref{DESY}}
\and B.~Khelifi \inst{\ref{APC}}
\and K.~Kosack \inst{\ref{IRFU}}
\and J.~Rico \inst{\ref{IFAE}}
\and A.~Sinha \inst{\ref{APC}}\thanks{\textit{$^{\star}$Send offprint requests to}: C.~Nigro, \texttt{cosimo.nigro@desy.de}}
}

\institute{ 
DESY, D-15738 Zeuthen, Germany \label{DESY} 
\and Max-Planck-Institut f\"ur Kernphysik, P.O. Box 103980, D 69029 Heidelberg, Germany \label{MPIK}
\and Landessternwarte, Universit\"{a}t Heidelberg, K\"{o}nigstuhl, D 69117 Heidelberg, Germany  \label{HDL}
\and Instituto de Astrof\'isica de Andaluc\'ia - CSIC, Glorieta de la Astronom\'ia s/n, 18008 Granada, Spain \label{IAA} 
\and Unidad de Part\'iculas y Cosmolog\'ia (UPARCOS), Universidad Complutense, E-28040 Madrid, Spain \label{UCM}
\and APC, AstroParticule et Cosmologie, Universit\'{e} Paris Diderot, CNRS/IN2P3, CEA/Irfu, Observatoire de Paris, Sorbonne Paris Cit\'{e}, 10, rue Alice Domon et L\'{e}onie Duquet, 75205 Paris Cedex 13, France \label{APC} 
\and TU Dortmund, Astroteilchenphysik E5b, 44227 Dortmund, Deutschland\label{DORTMUND}
\and Department of Physics and Astronomy, University of California, Los Angeles, CA 90095, USA \label{UCLA}
\and Physics Department, McGill University, Montreal, QC H3A 2T8, Canada \label{MCGILL}
\and Institut de Astrof\'{i}sica d’Altes Energies (IFAE), The Barcelona Institute of Science and Technology (BIST), E-08193 Bellaterra (Barcelona), Spain \label{IFAE} 
\and LUTH, Observatoire de Paris, PSL Research University, CNRS, Universit\'e Paris Diderot, 5 Place Jules Janssen, 92190 Meudon, France \label{LUTH} 
\and IRFU, CEA, Universit\'e Paris-Saclay, F-91191 Gif-sur-Yvette, France \label{IRFU}
}



\makeatletter
\renewcommand*{\@fnsymbol}[1]{\ifcase#1\@arabic{#1}\fi}
\makeatother

\abstract{The analysis and combination of data from different gamma-ray instruments involves the use of collaboration proprietary software and case-by-case methods. The effort of defining a common data format for high-level data, namely event lists and instrument response functions (IRFs), has recently started for very-high-energy gamma-ray instruments, driven by the upcoming Cherenkov Telescope Array (\cta). In this work we implemented this prototypical data format for a small set of \magic, \veritas, \fact, and \hess Crab nebula observations, and we analyzed them with the open-source \gammapy software package. By combining data from \fermi, and from four of the currently operating imaging atmospheric Cherenkov telescopes, we produced a joint maximum likelihood fit of the Crab nebula spectrum. Aspects of the statistical errors and the evaluation of systematic uncertainty are also commented upon, along with the release format of spectral measurements. The results presented in this work are obtained using open-access on-line assets that allow for a long-term reproducibility of the results.}


\maketitle

\section{Introduction}
The opening of new astronomical windows at different wavelengths in the last decades has made evident that many astrophysical puzzles could be solved only by combining images obtained by different facilities. In the late 1970s a common format was developed to facilitate the image interchange between observatories, hence overcoming incompatibilities between the numerous operating systems. The Flexible Image Transport System (FITS) format was standardized in 1980 \citep{Wells_1981} and formally endorsed in 1982 by the International Astronomical Union (IAU), that in 1988 formed a FITS working group (IAUFWG) entrusted to maintain the format and review future extensions. In the mid-1990s the NASA high energy astrophysics science archive research center (HEASARC) FITS Working Group, also known as the OGIP (Office of Guest Investigator Programs), promoted multi-mission standards for the format of FITS data files in high-energy astrophysics and produced a number of documents and recommendations that were subsequently incorporated into the FITS standard format definition. Since its conception, the FITS format has been been updated regularly to address new types of metadata conventions, the diversity of research projects and data product types. The last version of the FITS Standard document (4.0) was released in 2018\footnote{\url{https://fits.gsfc.nasa.gov/fits_standard.html}}. 
\par
Today the FITS format is in widespread use among astronomers of all observing bands, from radio frequencies to gamma rays. For instance, the high-energy gamma-ray (HE, E\SI{>100}{\mev}) Large Area Telescope (LAT, \citealt{fermi_tech}), on board of the \textit{Fermi} satellite, publicly releases all its high-level analysis data in FITS format, that, processed with the science tools, are used to obtain scientific products as spectra, light-curves and sky-maps. However, as a branch of astroparticle physics, very-high-energy (VHE, E$>$100\,GeV) gamma-ray astronomy inherited its methodologies and standards from particle physics, where the ROOT\footnote{\url{https://root.cern.ch/}} framework \citep{brun_1997} and its associated file format is commonly used. Despite the common container format neither the internal data structure nor the software is shared among the different experiments. VHE gamma-ray astronomy is conducted by ground-based telescopes, with the Imaging atmospheric Cherenkov telescopes (IACTs) among the most successful ones \citep{Naruois_2015}. Data from four of the currently operating IACTs were used in this project: the Major Atmospheric Gamma Imaging Cherenkov Telescopes (\magic, \citealt{MAGIC_2014_hardware}), the Very Energetic Radiation Imaging Telescope Array System (\veritas, \citealt{VERITAS_2006}), the First G-APD Cherenkov Telescope (\fact, \citealt{anderhub2013design}) and the High Energy Stereoscopic System (\hess, \citealt{hess_status}). Each of them is described in section~\ref{sec:data}.
\par
A new era in VHE gamma-ray astronomy is expected to start with the future Cherenkov Telescope Array (\cta, \citealt{CTA}), the next generation IACT instrument, which is currently under construction. The future operation of \cta as an observatory calls for its data format and analysis software to be available to a wide astronomical community. This requirement led to a standardization of the IACT data format, adopting the FITS standard, and the development of open-source science tools, initiating the integration of the VHE discipline into multi-instrument astronomy. A first attempt to define a common data format for the VHE gamma-ray data is being carried out within the ``Data formats for gamma-ray astronomy''\footnote{\gadfUrl} forum \citep{Deil_2016,v02}. This is a community effort clustering together members of different IACT collaborations with \cta as driving force. In this paper we implement this prototypical data format for data samples by \magic, \veritas, \fact, and \hess and we combine, for the first time, observations by \fermi and these four IACTs relying for the scientific analysis solely on open-source software, in particular on the \gammapy\footnote{\gammapyUrl} project \citep{2015ICRC...34..789D,Deil_2017}. We provide the reader not only with the datasets but also with all the scripts and an interactive environment to reproduce all the results. They are available at \url{https://github.com/open-gamma-ray-astro/joint-crab} and will be referred to, from now on, as online material. This allows to fully reproduce the results presented in the paper. 
The Crab nebula is selected as a target source for this work, being the reference source in the VHE gamma-ray astronomy \citep{2004ApJ...614..897A, HESS_crab, 2008ApJ...674.1037A, performance_stereo_MAGIC, MAGIC_2014_software} due to its brightness, apparent flux steadiness and visibility from all the considered observatories. 
\par 
The paper is structured as follows: we describe the selected datasets in Section~\ref{sec:data}, the process of extracting the spectral information with some considerations on the handling of statistical and systematic uncertainties in Section~\ref{sec:analysis}. In Section~\ref{sec:github} we present the resources we use to ensure the analysis reproducibility, and in Section~\ref{sec:extens} we provide some prospects for the reuse of the methodologies of data release discussed and implemented in this work. 
This report is a technical paper to show the ease of multi-instrument results once the format standardization is reached: we do not seek to draw any scientific conclusion on the physics of the Crab nebula, and of pulsar wind nebulae, in general.

\section{Datasets} \label{sec:data}
Differently than other astronomical telescopes, instruments for gamma astronomy cannot directly scatter or reflect gamma rays, being photon-matter interactions dominated by pair production for $E_{\gamma} > \SI{30}{MeV}$. The experimental techniques, either space-borne or ground-based \citep{funk_review}, rely on the direct detection of secondary charged particles or on the indirect detection of the Cherenkov emission of a cascade of charged secondaries they produce in the atmosphere. A detection, or event, cannot be unambiguously discriminated from the irreducible charged cosmic ray background, but can only be classified with a certain probability as a primary photon. Gamma-ray astronomy could therefore be labelled as ``event-based'' in contrast to the ``image-based'' astronomy where charge-coupled devices (CCDs) act as detectors. The input for the high-level analysis of gamma-ray astronomy data is typically constituted by two elements. The first one is a list of events that are classified (according to selection cuts) as gamma rays, along with their estimated direction, $\bm{P'}$, estimated energy, $E'$, and arrival time.
The second element consists of the instrument response functions (IRFs), quantifying the performance of the detector and connecting estimated quantities ($E'$, $\bm{P'}$) with their true, physical, values ($E$, $\bm{P}$). IRFs components are:
\begin{itemize}
\item effective area, representing the effective collection area of the instrument, $A_{\rm eff}(E, \bm{P})$; 
\item energy dispersion, the probability density function of the energy estimator $f_{E}(E'|E, \bm{P})$; 
\item point spread function (PSF), the spatial probability distribution of the estimated event directions for a point source, $f_{\bm{P}}(\bm{P'}|E, \bm{P})$. 
\end{itemize}
Their formal definition is shared with lower-energy instruments (e.g. x-ray, \citealt{x_ray_irfs}) and they are computed from Monte Carlo simulations. Since the detector response is not uniform across the field of view (FoV), the IRFs generally depend on the radial offset from the FoV center (full-enclosure IRFs). When this dependency is not taken into account, and a cut on the direction offset of the simulated events is applied, the IRFs are suited only for the analysis of a point-like source sitting at an a priori defined position in the FoV (point-like IRFs). IRFs components, in this case, do not have a dependency on the event coordinate, $\bm{P}$, but only on the energy and the PSF component is not specified. The differences between full-enclosure and point-like IRFs are illustrated in the interactive notebook \texttt{1\_data.ipynb} in the online material. For this publication we will use all the datasets to perform a point-like analysis.
In the IACT terminology, event lists and IRFs are dubbed Data Level\,3 (DL3) products \citep{CTA_data_spec}. The datasets used in this work include observations of the Crab nebula by \fermi, \magic, \veritas, \fact, and \hess following the format specifications available in the ``Data formats for gamma-ray astronomy'' forum \citep{Deil_2016}. The IACT DL3 datasets were produced with proprietary codes that extracted the event lists, and IRFs, and saved them in the requested format\footnote{All the FACT software used to generate DL3 datasets is open-source.}. They are released in chunks, typically of 20-30 minutes, of data acquisition, named runs. IACTs are ground-based, pointing instruments and their response varies with the observing conditions (atmospheric transmission, zenith angle, night sky background level, position of the source in the FoV) therefore their data come with per-run IRFs. The \fermi telescope, orbiting around the Earth at $\sim \SI{600}{km}$, is generally operating in survey mode and has a stable set of IRFs shipped with the science tools. We use the \fermi science tools and \texttt{gammapy} to make these datasets spec-compliant. This work constitutes the first joint release of datasets from different instruments in VHE gamma-ray astronomy. \newline
An interactive notebook illustrating the content of the datasets, \texttt{1\_data.ipynb}, is available in the online material. The datasets are presented in what follows in order of increasing instrument energy threshold (see Table~\ref{tab:datasets}). 

\begin{table}[htbp]
\caption{Crab nebula datasets used in this work. $T_{\rm obs}$ stands for observation time, $E_{\rm min}$ and $E_{\rm max}$ identify the energy range of the analysis, $N_{\rm on}$ and $N_{\rm bkg}$ the number of total and background events, respectively, estimated in the circular signal region with a radius $R_{\rm on}$. 
\label{tab:datasets}}
\centering
\begin{tabular}{lrrrrrl}
\hline
Dataset &  $T_{\rm obs}$ & $E_{\rm min}$ & $E_{\rm max}$ & $N_{\rm on}$ & $N_{\rm bkg}$ & $R_{\rm on}$  \\
        &       & TeV           & TeV           &              &         & deg        \\ \hline
\fermi & $\sim$7 yr & 0.03 & 2 & 578 & 1.2 & 0.30 \\
\magic & 0.66 h & 0.08 & 30 & 784 & 129.9 & 0.14 \\
\veritas & 0.67 h & 0.16 & 30 & 289 & 13.7 & 0.10 \\
\fact & 10.33 h & 0.45 & 30 & 691 & 272.8 & 0.17 \\
\hess & 1.87 h & 0.71 & 30 & 459 & 27.5 & 0.11 \\
\hline
\end{tabular}
\end{table}

\begin{figure}
	\centering
	\includegraphics[scale=0.5]{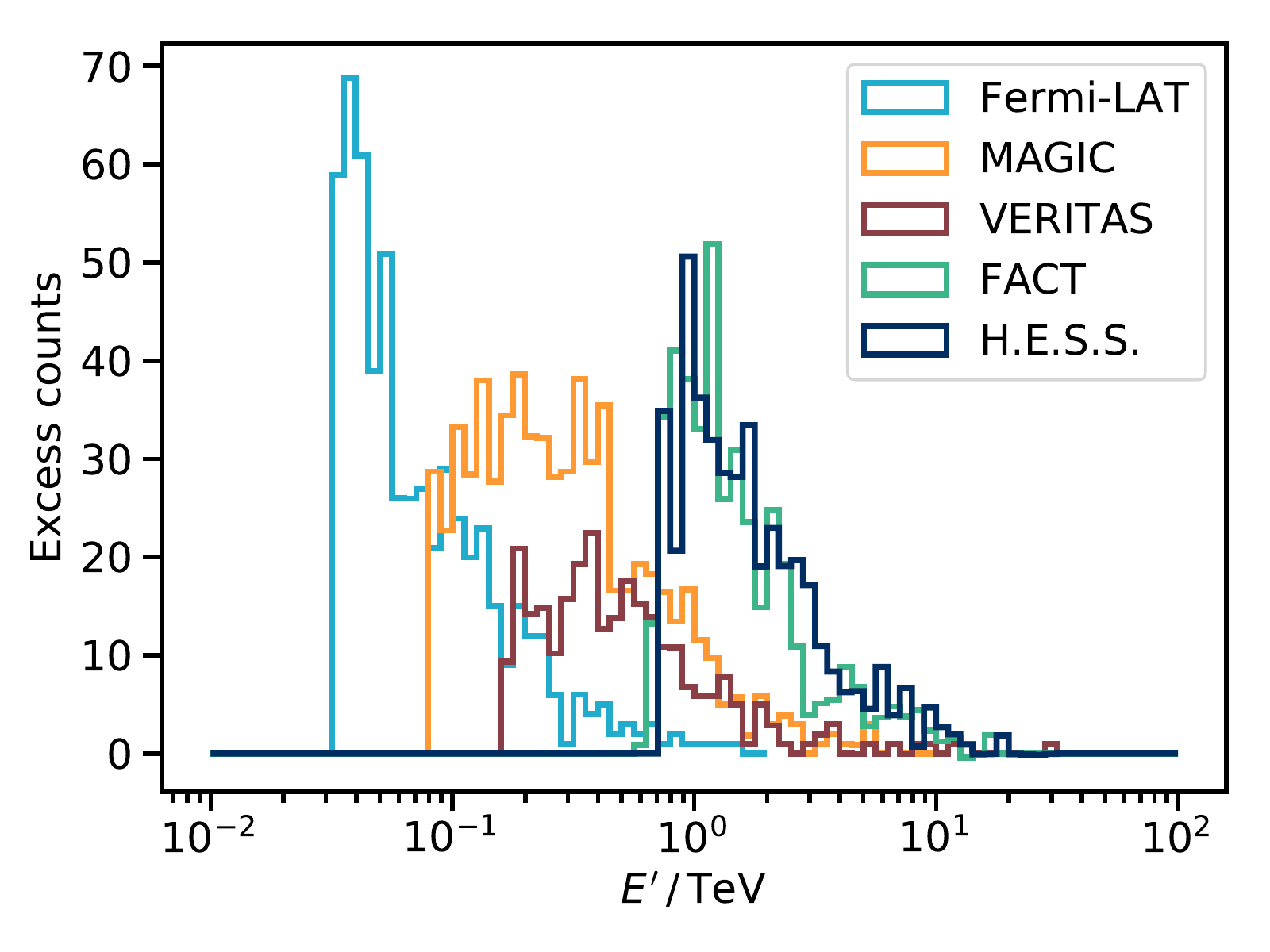}
	\caption{Histograms of the estimated mean number of signal events from the Crab nebula, number of excess events, vs estimated energy for each dataset.}
	\label{fig:count_spec}%
\end{figure}

\subsection{Fermi-LAT} 
\label{sec:fermi}
The LAT detector, on board of the \textit{Fermi} satellite, is an imaging pair-conversion telescope, which has been designed to detect photons between \SI{20}{MeV} and more than \SI{300}{GeV} \citep{fermi_tech}. We analyzed the publicly available observations spanning from 2008 August 8 to 2015 August 2, i.e. $\sim$7\,years of operations, in a 30$^{\circ}$-radius region around the position of the Crab nebula. We used all FRONT and BACK type events belonging to the Source class with a reconstructed direction within 105$\deg$ from the local zenith (to reject the emission from the Earth's atmosphere) and a reconstructed energy between \SI{30}{GeV} and \SI{2}{TeV}. The lower energy cut was chosen to minimize the contamination of the Crab pulsar that is located at the center of the nebula. We estimated that above \SI{30}{GeV} the pulsar emission contributes to less than 10\% to the detected flux and can be neglected given the technical purpose of the paper. An interactive notebook illustrating this calculation is available in the online material (\texttt{5\_crab\_pulsar\_nebula\_sed.ipynb}).
To reduce the IRFs to a DL3-compliant format, we compute the PSF using \texttt{gtpsf} and estimated the effective area from the sky-coordinate and energy-dependent exposure computed with \texttt{gtexpcube2}. The effective area is simply the exposure scaled by the observation time. The energy dispersion at the energies considered in this analysis has an impact smaller than 5\% on the reconstructed spectrum\footnote{\url{https://fermi.gsfc.nasa.gov/ssc/data/analysis/documentation/Pass8\_edisp\_usage.html}}. We approximate the energy dispersion with a Gaussian distribution with mean (bias) 0 and standard deviation 0.05 (resolution) independent on the estimated direction and energy. Event list and IRFs produced by the \fermi science tools are already FITS files, \texttt{gammapy} is used to store them in a DL3-compliant format. 

\subsection{\magic}
\magic is a system of two 17-m diameter IACTs, with $3.5^{\circ}$ FoV located on the Canary island of La Palma, Spain at the Roque de Los Muchachos Observatory (2200\,m above sea level). The first telescope worked in stand-alone mode from 2004 until 2009, when a second one started operations. In 2012 the two MAGIC telescopes underwent a major upgrade of the readout systems and the camera of the first telescope \citep{MAGIC_2014_hardware} leading to a significant improvement of the instrument performance \citep{MAGIC_2014_software}. The FITS data released by the MAGIC collaboration includes 40~min of Crab nebula observations and their corresponding IRFs. The data were recorded in 2013, after the aforementioned major upgrade, at small zenith angles ($<30^{\circ}$) in wobble mode \citep{wobble} with the source sitting at an offset of $0.4^{\circ}$ from the FoV center. They are chosen from the data sample used for the performance evaluation in \citet{MAGIC_2014_software}. The IRFs released by the \magic collaboration were generated using the MARS software \citep{MAGIC_Mars}, for a point-like source response: i.e. they are calculated for a simulated source at $0.4^{\circ}$ offset, applying a directional cut on the events direction of $0.14^{\circ}$ around the source position.

\subsection{\veritas}
\veritas consists of four 12-m diameter IACTs \citep{VERITAS_2006}, with $3.5^{\circ}$ FoV based at the Fred Lawrence Whipple Observatory in Southern Arizona, USA. Since the start of full array operations in 2007, the sensitivity of VERITAS was improved by two major upgrades: the  relocation  of one of its telescopes in 2009 \citep{VERITAS_relocation} and the upgrade of the camera with higher quantum efficiency photomultiplier tubes in 2012 \citep{VERITAS_upgrade_2}. The DL3 data released by the VERITAS collaboration includes 40~min of archival observations of the Crab nebula taken in 2011, after the telescope relocation, but prior to the camera upgrade. These observations were carried out in wobble mode with the standard offset of 0.5$^{\circ}$ at small zenith angles $(<20^{\circ})$. The released IRFs are valid for the analysis of point-like sources taken with the standard offset angle and were generated from events surviving the standard directional cut of 0.1$^{\circ}$, using the VEGAS software package \citep{VERITAS_VEGAS}. 

\subsection{\fact}
\fact\citep{anderhub2013design, fact_performance} is a single IACT with 4-m diameter reflective surface and a FoV of 4.5$^{\circ}$, that is located next to MAGIC at the Roque de Los Muchachos Observatory. \fact tests the feasibility of silicon photo-multipliers (SiPM) for use in VHE gamma-ray astronomy. It is the first fully automated Cherenkov telescope that takes data without an operator on site \citep{fact_automatic}. For this work \fact released one full week of observations of the Crab nebula taken in November 2013, corresponding to \SI{10.3}{\hour}\footnote{\url{https://factdata.app.tu-dortmund.de/}}. The data were recorded in wobble mode with an offset angle of $0.6^{\circ}$ at zenith angles smaller than $30^{\circ}$. The corresponding IRFs are of point-like type with a directional cut of $0.17^{\circ}$. 

\subsection{\hess}\label{sec:hess}
\hess is an array of five IACTs located in Namibia, on the Khomas Highland, near the Gamsberg mountain. The first four 12-m diameter telescopes, arranged in a square, became operational in December 2003 marking the start of what today is called \hessOne with a FoV of $5^{\circ}$. Since July 2012 a fifth 28-m diameter telescope, located at the array center, started operation (\hessTwo) both in stereoscopic and stand-alone mode. In this work, we used four observation runs of the Crab nebula carried out by \hessOne in 2004, each of them with a duration of 28~min. They were taken in wobble mode at zenith angles between 45$^{\circ}$ and 50$^{\circ}$, half of them with a 0.5$^{\circ}$, and the other half with a 1.5$^{\circ}$ offset angle.  These data are the Crab runs part of the first FITS test data release \citep{hess_dl3}\footnote{\hessDataUrl}. \hess released full-enclosure IRFs \citep{Deil_2016}, i.e. no directional cut is applied on the simulated events. 

\bigskip
\par
\section{Data analysis} \label{sec:analysis}
In this section we present a spectral analysis of the gamma-ray datasets described in Section~\ref{sec:data}. First, the gamma-ray event data and IRFs are reduced for each instrument (Section~\ref{sec:an_sp}). Then, in Section~\ref{sec:likelihood}, we preform a spectral likelihood fit, under the assumption of a log-parabola analytic model, for each datasets separately, and for all the datasets together (joint fit). Finally, we present an analysis that includes a systematic error term, representing the uncertainty on the energy scale of each instrument, in a modified likelihood function. 

\subsection{Spectrum extraction}\label{sec:an_sp}
In order to estimate the energy spectrum of a gamma-ray source ($\frac{{\rm d}\phi}{{\rm d}E}(E;\vect{\Lambda})$, with  $\vect{\Lambda}$ a the set of spectral parameters), a binned maximum likelihood method, with $n_{E'}$ bins in estimated energy $E'$ is used. The observed data $\vect{D}$ for such a likelihood function are the number of events in a circular signal region (labeled as ON) containing the gamma-ray source and in a control region (labeled as OFF) measuring the background to be subtracted from the ON. Considering  $n_{\rm runs}$ observation runs from $n_{\rm instr}$ different instruments (or datasets), we can write the likelihood as:
\begin{equation}
\mathcal{L}(\vect{\Lambda} | \vect{D})= \\
\prod^{n_{\rm instr}}_{i=1} \mathcal{L}_i(\vect{\Lambda} | \{ N_{\mathrm{on}, ijk}, N_{\mathrm{off}, ijk} \}_{j=1,...,n_{\rm runs}; k=1,...,n_{E'}})
\label{eq:joint_likelihood}
\end{equation}
with each instrument contributing with a term:
\begin{equation}
\begin{split}
& \quad
\mathcal{L}_i(\vect{\Lambda} | \{ N_{\mathrm{on}, ijk}, N_{\mathrm{off}, ijk} \}_{j=1,...,n_{\rm runs}; k=1,...,n_{E'}}) = \\
& \quad
\prod^{n_{\rm runs}}_{j=1} \prod^{n_{E'}}_{k=1} 
\operatorname{Pois}(g_{ijk}(\vect{\Lambda}) + b_{ijk}; N_{\mathrm{on}, ijk}) \times \operatorname{Pois}(b_{ijk}/\alpha_{ij}; N_{\mathrm{off}, ijk}),
\end{split}
\label{eq:joint_likelihood_instr_term}
\end{equation}
where:
\begin{itemize}
\item $N_{\mathrm{on}, ijk}$ and $N_{\mathrm{off}, ijk}$ are the number of observed events within the ON and OFF regions, respectively, in the energy bin $k$ in the run $j$ for the $i$-th instrument. They are both characterized by a Poisson distribution;
\item $g_{ijk}(\vect{\Lambda})$ and $b_{ijk}$ are the expected number of signal and background events, respectively, in the energy bin $k$ in the run $j$ for the $i$-th instrument. $g_{ijk}$ is computed with the forward folding technique: for a point-like analysis the assumed spectrum $\frac{{\rm d} \phi}{{\rm d} E}$ is convolved with the effective area and energy dispersion IRFs component. $b_{ijk}$, in absence of a background spectral model, is treated as a nuisance parameter and fixed to the value returning $\frac{\partial \mathcal{L}}{\partial b_{ijk}} = 0$, for the mathematical details of $g_{ijk}$ and $b_{ijk}$ evaluation see Appendix A in \citet{piron}. 
\item $\alpha_{ij}$ is the ON to OFF exposures ratio, constant with energy in our case, in the run $j$ for the $i$-th instrument.
\end{itemize}
The size of the circular ON region per each dataset is given as $R_{\rm on}$ in \mbox{Table~\ref{tab:datasets}}, the OFF region can be defined either as multiple regions mirroring the ON symmetrically with respect to the telescope pointing position (reflected regions background \citealt{2001A&A...370..112A,Berge2007}) or as a ring around the source position (ring background method in \citealt{Berge2007}). Given the wobble mode observation strategy, and the small FoV, the reflected regions method is naturally suitable for the IACTs datasets. On the other hand, the ring background method is used for the \fermi datasets, in this analysis with a circular signal region of 0.3$^{\circ}$ radius and a background ring with inner and outer radius of $1^{\circ}$ and $2^{\circ}$, respectively. We choose, for all the instruments, 20 bins per decade for the estimated energy between $10\,{\rm GeV}$ and $100\,{\rm TeV}$. For a given instrument $i$ and run $j$ the likelihood values are not computed in the energy bins outside the range $\left[E_{\rm min}, E_{\rm max}\right]$ given in \mbox{Table~\ref{tab:datasets}}. The choice of the energy range for \fermi is already discussed in Section~\ref{sec:fermi}. For the IACT datasets, $E_{\rm min}$ is a safe energy threshold for the spectrum extraction computed by the collaboration software and hard coded in the DL3 files. It is mostly dependent on the experiment performance and on the zenith angle of the observations. \fact has an energy threshold of $450$\,GeV, higher than MAGIC and VERITAS despite the observations carried out in the same zenith angle range, due to its limited light-collection area and the single telescope observations. The larger zenith angle of the \hess datasets is due to the low altitude at which the source culminates at the \hess site. This yields an energy threshold of $\sim 700$\,GeV, higher than any other IACT. The maximum energy $E_{\rm max}$ is fixed to 30\,TeV for all the IACTs and it is chosen to cover the whole energy range containing events. 
\newline
The mean number of signal events, labeled as excess events, can be estimated via the equation $N_{\mathrm{ex}, ijk} = N_{\mathrm{on}, ijk} - \alpha_{ij} N_{\mathrm{off}, ijk}$. The distribution of the excess events in each estimated energy bin, summed over all the observational runs per each instrument $(\sum^{n_{\rm runs}}_{j=1} N_{\mathrm{ex}, ijk})$ is shown in \mbox{Figure~\ref{fig:count_spec}}. \mbox{Table~\ref{tab:datasets}} reports also the total number of observed events in the ON region $(N_{\rm ON} = \sum_{ijk} N_{\mathrm{on}, ijk})$ per each experiment, and the number of background events in the ON region per each experiment, obtained scaling the OFF events with the exposures ratio $\alpha_{ij}$ $(N_{\rm bkg} = \sum_{ijk} \alpha_{ij} N_{\mathrm{off}, ijk})$. 
\par
\noindent
To perform a joint point-like analysis we reduce the \fermi and \hess full-enclosure IRFs to a point-like format, removing the dependency from the source position in the FoV. For \fermi, under the assumption that the acceptance is uniform in a small sky region close to our target, we obtained a point-like effective area by taking the value at the source position. In each energy bin we corrected the effective area with a containment fraction computed integrating the PSF over the signal region. Similarly for the \hess IRFs the value at the source offset is taken and a correction based on the PSF containment fraction is computed.

\begin{figure}
	\centering
	\includegraphics[width=\columnwidth]{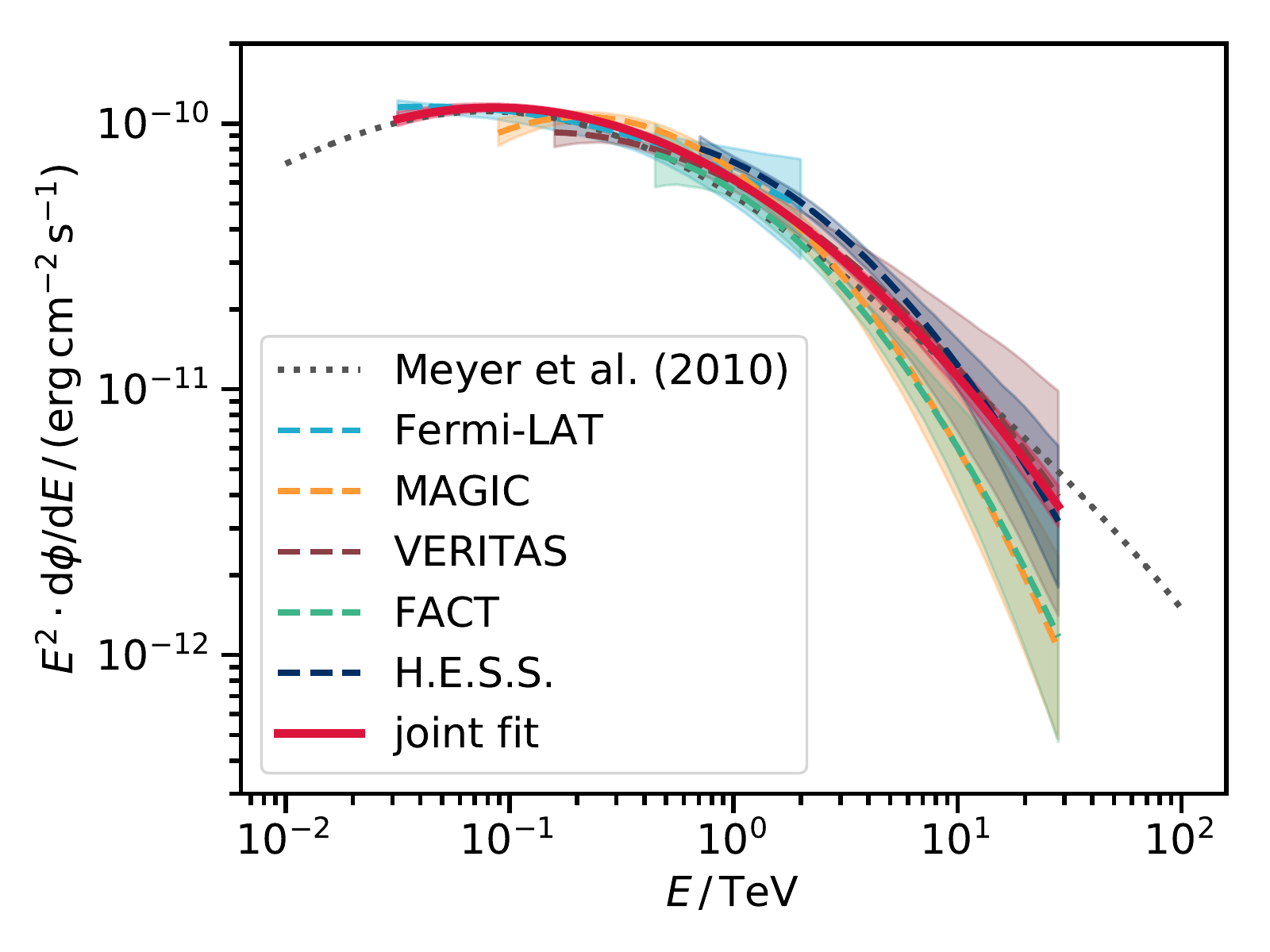}
	\caption{Crab nebula SED for individual instrument fits and from the joint fit. Single-instrument results are represented with dashed lines, the fit of all the datasets together, labeled as joint, is represented as a thick, solid red line. The shaded areas represent the SED error bands whose calculation is explained in Section~\ref{sec:likelihood}. The dotted line shows the model in \citet{Meyer2010}.}
	\label{Fig:joint_SED}%
\end{figure}

\begin{figure*}
	\centering
	\includegraphics[scale=0.45]{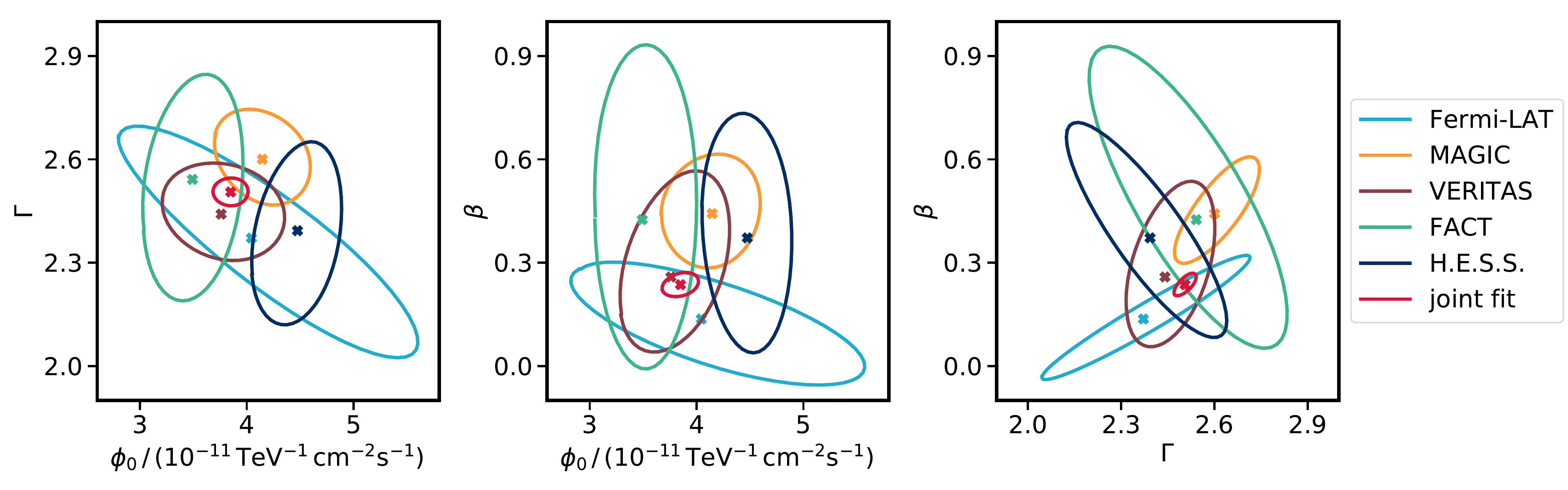}
	\caption{Likelihood contours corresponding to 68 \% probability content for the fitted spectral parameters $(\phi_0, \Gamma, \beta)$, for the likelihood in Eq.~\ref{eq:joint_likelihood}. Results from  the individual instruments and from the joint-fit are displayed.}
	\label{Fig:LP_contours}%
\end{figure*}

\subsection{Spectral model fit}
\label{sec:likelihood}
We assumed a spectral model of log-parabolic form:
\begin{equation}
\frac{{\rm d}\phi}{{\rm d}E} = \phi_0 \left( \frac{E}{E_0} \right)^{
    - \Gamma - \beta \log_{10}{ \left( \frac{E}{E_0} \right),
}
\label{eq:log-parabola}
}
\end{equation}
since it was suggested as  the best-function approximation for the Crab nebula spectrum in such a wide energy range \citep{MAGIC_2015_crab}.
The spectral parameters $\vect{\Lambda} = (\phi_0, \Gamma, \beta)$ are left free to vary in the fit while the reference energy $E_0$ is fixed at the value of 1\,TeV. We refer to the result of the maximum likelihood using all the instrument datasets as \textit{joint fit}. As a consistency check we also fitted each instrument dataset separately ($i$ fixed in \mbox{Eq.~\ref{eq:joint_likelihood}}). $E_0$ is usually chosen to minimize the correlation between the other spectral parameters. In this work, in order to directly compare the parameters $\vect{\Lambda}$ also for the fit with the individual instrument datasets, the reference energy $E_0$ is kept fixed at the same value of 1\,TeV. This introduces larger errors and correlation for the datasets for which such value is close to one of the extremes of its energy range. 
The resulting spectral energy distributions (SEDs, $E^2 {\rm d} \phi / {\rm d} E$) are shown in \mbox{Fig.~\ref{Fig:joint_SED}}, together with a theoretical model taken from \citet{Meyer2010}. The values of the fit parameters are listed in \mbox{Table~\ref{table:fit_results}}. The joint fit inherently comes with an increase in statistical power, as evidenced by the shrinking of the confidence contours of the fitted spectral parameters for the joint fit in \mbox{Fig.~\ref{Fig:LP_contours}}. We note that the Crab SED shape is not exactly represented by log parabola across the 30 GeV to 20 TeV energy range, which is one reason for differences in the measured fit parameters from the different experiments. An interactive summary of the spectral results is available in the online material (\texttt{2\_results.ipynb}).

\begin{table}[htbp]
\caption{Spectral model best-fit parameters, as defined in Eq.~\ref{eq:log-parabola}. The reference energy, $E_0$, is taken at 1\,TeV for all the fit results. The prefactor $\phi_0$ is given in $10^{-11}$\,TeV$^{-1}$\,cm$^{-2}$\,s$^{-1}$. 
\label{table:fit_results}}
\centering
\begin{tabular}{lccc}
\hline
Dataset &  $\phi_0$ & $\Gamma$ & $\beta$ \\
\hline
\fermi & 4.04 $\pm$ 1.01 & 2.37 $\pm$ 0.24 & 0.14 $\pm$ 0.13 \\
\magic & 4.15 $\pm$ 0.30 & 2.60 $\pm$ 0.10 & 0.44 $\pm$ 0.11 \\
\veritas & 3.76 $\pm$ 0.36 & 2.44 $\pm$ 0.09 & 0.26 $\pm$ 0.17 \\
\fact & 3.49 $\pm$ 0.30 & 2.54 $\pm$ 0.22 & 0.42 $\pm$ 0.31 \\
\hess & 4.47 $\pm$ 0.29 & 2.39 $\pm$ 0.18 & 0.37 $\pm$ 0.22 \\
joint & 3.85 $\pm$ 0.11 & 2.51 $\pm$ 0.03 & 0.24 $\pm$ 0.02 \\
\hline
\end{tabular}
\end{table}
\par

\begin{figure}[ht]
	\centering
	\includegraphics[width=\columnwidth]{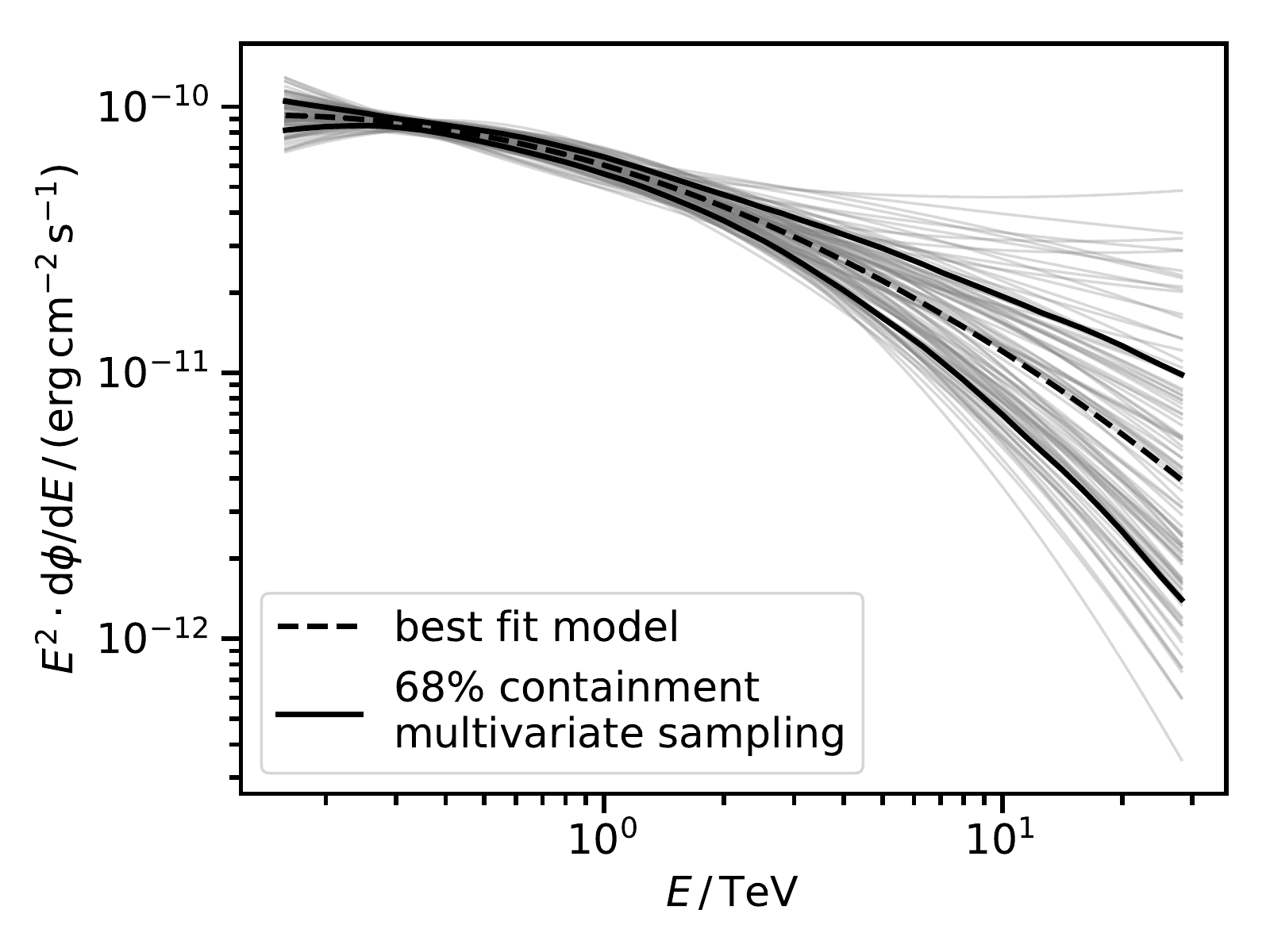}
	\caption{Error estimation methods for the measured SED using the \veritas dataset, as example case. The solid black lines display the upper and lower limits of the error band estimated with the multivariate sampling. They represent the 68\% containment of 500 spectral realizations (100 displayed as gray lines) whose parameters are sampled from a multivariate distribution defined by the fit results.
	\label{Fig:err_prop}}
\end{figure}

The statistical uncertainty on the SED is estimated by using a sampling technique to propagate the errors from the fit parameters. We assume that the likelihood of the model parameters $\vect{\Lambda}$ is distributed according to a multivariate normal distribution with mean vector $\vect{\mu}$ and covariance matrix $\vect{\Sigma}$ defined by the fit results. We assume $\vect{\mu} = \hat{\vect{\Lambda}} = (\hat{\phi_0}, \hat{\Gamma}, \hat{\beta})$, values of the fitted parameters and $\vect{\Sigma} = \hat{\vect{V}}_{\hat{\vect{\Lambda}}}$, covariance of the fitted parameters. We sample this distribution and compute the spectrum realization corresponding to each sampled $\vect{\Lambda}$. The $\pm 1\,\sigma$ uncertainty on the fitted spectrum is estimated by taking, at a given energy, the quantiles of the fluxes distribution that returns a 68\% containment of all the realizations. The upper and lower limits of the error band estimated with our method are plotted in black against 100 realizations of the spectrum with sampled parameters in \mbox{Fig.~\ref{Fig:err_prop}}, for the example case of the VERITAS datasets.

\subsection{Systematic uncertainties on different energy scales}
\begin{figure*}
\centering
\includegraphics[scale=0.45]{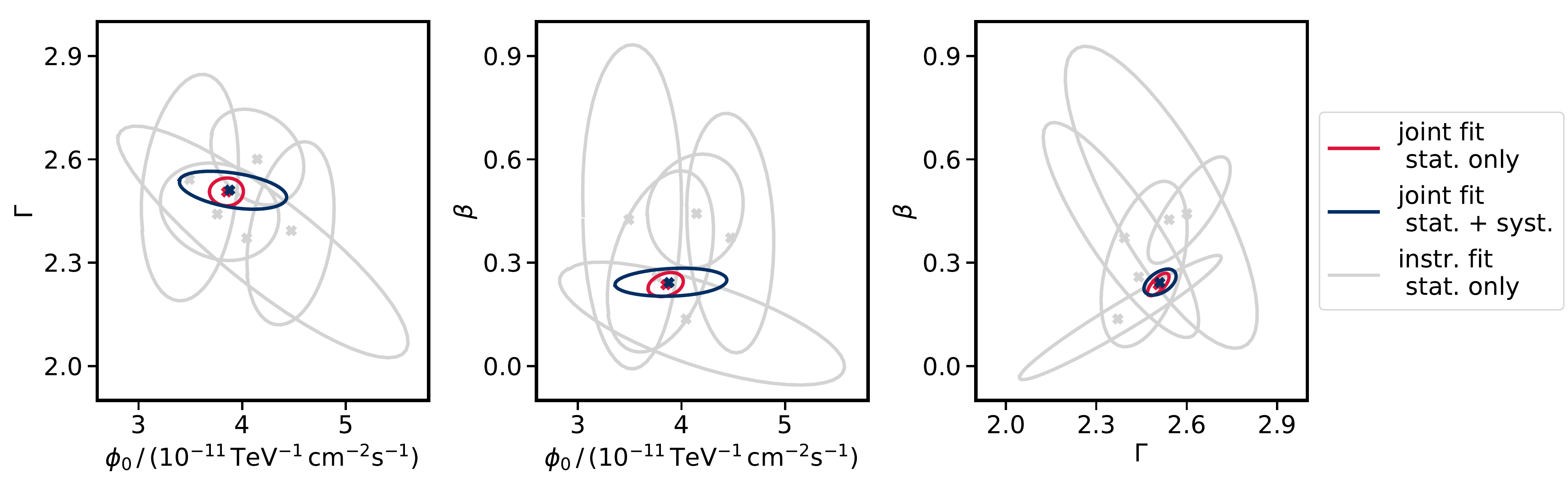}
\caption{Likelihood contours corresponding to 68 \% probability content for the fitted spectral parameters $(\phi_0, \Gamma, \beta)$, for the likelihood in Eq.~\ref{eq:joint_likelihood} (red) and the likelihood in Eq.~\ref{eq:joint_likelihood_syst} (blue). Results from the individual instruments with the likelihood in Eq.~\ref{eq:joint_likelihood} are displayed in gray.}
\label{fig:contours_syst}
\end{figure*}

Spectral measurements in gamma-ray astronomy are affected by multiple sources of systematic uncertainty. The DL3 data contains systematic uncertainties  that originate from an imperfect modeling of the atmosphere, telescopes and event reconstruction, resulting in a shift of the reconstructed energy scale and errors in the assumed IRF shapes. A second source of systematic error comes from the data reduction and model fitting, e.g. due to energy binning and interpolation effects, as well as from source morphology and spectral shape assumptions.
Generally two approaches are used to evaluate and report systematic errors \citep{2003PhRvD..67a2002C,2017arXiv170103701B}: multiple analyses / bracketing as in \citep{HESS_crab,performance_stereo_MAGIC} or modified likelihood with nuisance parameters \citep{Dickinson,dembinski,4FGL}.
The first approach leads to the estimation of an overall systematic error on each spectral parameter, e.g. for the flux normalization $\phi_0 \pm \sigma_{\phi_0, \mathrm{stat.}} \pm \sigma_{\phi_0, \mathrm{syst.}}$ , whereas the second method yields a global error including both statistical and systematic uncertainties, i.e. $\phi_0 \pm \sigma_{\phi_0, \mathrm{stat. + syst.}}$.
As an example of how to treat systematic errors, we present here an analysis with a modified likelihood that includes the uncertainty on the energy scale. 
Following \citet{dembinski}, we define a new joint likelihood function that includes a constant relative bias of the energy estimator per each instrument $z_i$, characterized by a Gaussian distribution with mean 0 and standard deviation $\delta_i$, the systematic uncertainty on the energy scale estimated by the single instrument ($\mathcal{N}(z_i; 0, \delta_i^2)$ in the following notation). This parameter is defined as $z_{i} = \frac{\tilde{E} - E}{E} = \frac{\tilde{E}}{E} - 1$, with $\tilde{E}$ being the energy reported by an instrument and $E$ the actual energy of each single event. The apparent spectral model we aim to fit for a single instrument would then be:
\begin{equation}
\frac{{\rm d}\tilde{\phi}}{{\rm d}\tilde{E}} = \frac{{\rm d}\phi}{{\rm d}E} \frac{{\rm d}E}{{\rm d}\tilde{E}} = \phi_0 
\left( \frac{E/(1+z)}{E_0} \right)^{-\Gamma + \beta \log_{10} \left( \frac{E/(1+z)}{E_0} \right)} \left( \frac{1}{1+z}\right)
\label{eq:log_par_energy_scale}
\end{equation}
and the overall joint likelihood is modified in:
\begin{equation}
\begin{split}
& \quad
\mathcal{L}(\vect{\Lambda} | \vect{D})= \\
& \quad
\prod^{n_{\rm instr}}_{i=1} \mathcal{L}_i(\vect{\Lambda} | \{ N_{\mathrm{on}, ijk}, N_{\mathrm{off}, ijk} \}_{j=1,...,n_{\rm runs}; k=1,...,n_{E'}}) \times \mathcal{N}(z_i; 0, \delta_i^2) 
\end{split}
\label{eq:joint_likelihood_syst}
\end{equation}
where now the energy biases are included in the spectral parameters to be fitted: $\vect{\Lambda} = (\phi_0, \Gamma, \beta, z_1, ..., z_{n_{\rm isntr}})$ and the energy spectrum ${\rm d}\tilde{\phi} / {\rm d}\tilde{E}$ in \mbox{Eq.~\ref{eq:log_par_energy_scale}} is used to predict the $g_{ijk}$ via forward folding. As in Eq.~\ref{eq:joint_likelihood}, $i$ runs over the instruments, $j$ on the runs and $k$ on the energy bin.
The inclusion of the energy biases allows, in addition to the variation of the global spectral parameters $\phi_0$, $\Gamma$ and $\beta$ (the same for all datasets), also an instrument-dependent energy adjustment (a shift) of the assumed model, through the individual $z_{i}$. This shift is not arbitrary: it is, in fact, constrained by its Gaussian distribution with a standard deviation given by the systematic uncertainty on the energy scale provided by the single experiment, $\delta_{i}$. Hereafter we will refer to this likelihood fit as \textit{stat.+syst. likelihood} that is  the generalized version of \mbox{Eq.~\ref{eq:joint_likelihood}} (obtainable from \mbox{Eq.~\ref{eq:joint_likelihood_syst}} simply fixing all $z_{i} = 0$). The result of the \textit{stat.+syst. likelihood} joint fit is shown in \mbox{Fig.~\ref{fig:contours_syst}} in blue against the result of the \textit{stat. likelihood} (\mbox{Eq.~\ref{eq:joint_likelihood}}) fit in red. 
We note that in this work we only account for the energy scale systematic uncertainty, as an example of a modified likelihood. A full treatment of the systematic uncertainty goes beyond the scope of this paper. It is possible to reproduce interactively the systematic fit in the online material (\texttt{3\_systematics.ipynb}).

\section{Reproducibility} \label{sec:github}
This work presents a first reproducible multi-instrument gamma-ray analysis, achieved by using the common DL3 data format and the open-source \texttt{gammapy} software package. We provide public access to the DL3 observational data, scripts used and obtained results with the GitHub repository mentioned in the introduction, along with a Docker container\footnote{\url{https://hub.docker.com/r/gammapy/joint-crab}} on DockerHub, and a Zenodo record \citep{joint-crab}, that provides a Digital Object Identifier (DOI). The user access to the repository hosting data and analysis scripts represents a necessary, but not sufficient condition to accomplish the exact reproducibility of the results. We deliver a \texttt{conda}\footnote{\url{https://conda.io}} configuration file to build a virtual computing environment, defined with a special care in order to address the internal dependencies among the versions of the software used. Furthermore, since the availability of all the external software dependencies is not assured in the future, we also provide a \texttt{joint-crab docker} container, to guarantee a mid-term preservation of the reproducibility. The main results published in this work may be reproduced executing the \texttt{make.py} command. This script works as a documented command line interface tool, wrapping a set of actions in different option commands that either extract or run the likelihood minimization or reproduce the figures presented in the paper. 
\par
The documentation is provided in the form of Jupyter notebooks. These notebooks can also be run through Binder\footnote{\url{https://mybinder.org/v2/gh/open-gamma-ray-astro/joint-crab/master?urlpath=lab/tree/joint-crab}} public service to access via web browser the whole \texttt{joint-crab} working execution environment in the Binder cloud infrastructure. The Zenodo joint-crab record, the joint-crab docker container, and the joint-crab working environment in Binder may be all synchronized if needed, with the content present in the joint-crab GitHub repository. Therefore, if eventual improved versions of the joint-crab bundle are needed (i.e. comments from referees, improved algorithms or analysis methods, etc.), they may be published in the GitHub repository and then propagated from GitHub to the other joint-crab repositories in Zenodo, DockerHub and Binder. All these versions would be kept synchronized in their respective repositories.

\section{Extensibility}
\label{sec:extens}
Another significant advantage of the common-format, open-source and reproducible approach we propose to the VHE gamma-ray community is the possibility to access the ON and OFF events distributions and the IRFs, i.e. the results of the spectrum extraction, saved in the OGIP spectral data format\footnote{\url{https://gamma-astro-data-formats.readthedocs.io/en/latest/spectra/ogip/index.html}} (they can be interactively accessed in \texttt{1\_data.ipynb}). This would allow to perform a maximum likelihood fit to any assumed spectral model ${\rm d}\phi/{\rm d}E$, that is otherwise impossible. This is of crucial interest for researchers not associated to experimental collaborations that, having usually access only to the final spectral points (often published with no covariance matrix attached), cannot properly test their theoretical models against the data. In the online material of this work, besides the analytically log-parabola function, we considered also a theoretical Synchrotron-Self Compton radiative model (\texttt{4\_naima.ipynb}) obtained with the open-source \texttt{naima} Python package \citep{naima}. This is meant to emphasize, on one hand, the potential of the proposed approach, and, on the other hand, the easy interchange between open-source astronomical Python packages with different functionality. 

\section{Conclusions}
This paper presents a multi-instrument reproducible gamma-ray analysis realized with open-source software. 
It also contains the first public joint release of data from IACTs. 
Such data dissemination offers the astronomical community the opportunity to gather knowledge of the VHE analysis techniques while waiting for the forthcoming \cta operations. Furthermore they can also be used in data challenges or coding sprints to improve the status of the current science tools.
\par
On a technical note, the DL3 data producible at the moment allow only for joint analyses of a target source at a given position of the FoV (the Crab nebula in this case): no other potential source in the FoV (or multiple sources) can be analysed given the point-like IRFs computed accordingly with the source position that all the instruments, but \hess, made available. It is worth noting that even the more exhaustive full-enclosure IRFs format may require further development as the current radial offset dependency does not account for a possible non-azimuthal symmetry of the instrument acceptance \citep{prandini_2015}.
\par
On a more general note, the objective of this publication is also to remark a novel approach to gamma-ray science, summarized through the three essential concepts of: common data-format, open-source software and reproducible results. 
We illustrate that a common data-format allows naturally multi-instrument analysis. Generating data samples compliant with the prototypical DL3 format defined in the ``Data formats for gamma-ray astronomy'' forum, we perform, for the first time, a spectral analysis of the Crab nebula using data from \fermi and four currently operating IACTs by using the \gammapy software package. Open-source software will be a key asset for the upcoming CTA, which, as an open-observatory, will share its observation time and data with the wider astronomical community.
Reproducible results are seamlessly achieved once the data and software are publicly available. There are several tools and platforms on the market that can be used for this purpose. In particular, with the on-line material attached to this issue, we show a practical example of how a future gamma-ray publication can be released with long-term solutions. A Git repository suffices for the first period after the publication, whereas a Docker accounts for the eventual loss of maintenance of the software packages needed for the analysis.    
We also provided some considerations on analysis procedures related to spectral analysis commonly performed in the VHE IACT-related astronomy. We proposed a method for computing error bands on the measured SED based on the sampling of a multivariate distribution, along with a method to account for the systematic uncertainties on the energy scales of different gamma-ray instruments while performing a joint fit of their data. We also pointed-out the advantages of publishing the outputs of the spectrum extraction, i.e. the distribution of the signal and background events and the IRFs (alike the OGIP spectral data in the \texttt{joint-crab} repository), instead of the spectral points. Mainly this grants the possibility to successively construct a likelihood using an arbitrary theoretical spectral model. 

\begin{acknowledgements}
This work was supported by the Young Investigators Program of the Helmholtz Association, by the Deutsche Forschungsgemeinschaft (DFG) within the Collaborative Research Center SFB 876 "Providing Information by Resource-Constrained Analysis", project C3 and by the European Commision through the ASTERICS Horizon2020 project (id 653477).  
We would like to thank the \hess, \magic, \veritas and \fact collaborations for releasing the data that were used. This work made use of \texttt{astropy} \citep{astropy} and \texttt{sherpa} \citep{sherpa}.
The authors are indebted to Abelardo Moralejo and Hans Dembinski for their useful suggestions on the statistical and systematic uncertainty estimation. We are grateful to the anonymous referee for improving the paper with his helpful comments.

\end{acknowledgements}

\bibliographystyle{aa} 
\bibliography{mybibfile}

\end{document}